\documentclass[10pt]{article}
\usepackage{cite} 
\usepackage{graphicx,amsmath,amssymb,amsbsy,latexsym,amsthm}
\usepackage[mathscr]{eucal}

\theoremstyle{plain}
\newtheorem{theorem}{Theorem}
\newtheorem{proposition}[theorem]{Proposition}
\newtheorem{lemma}[theorem]{Lemma}

\newcommand{\startproof}{\setlength{\parindent}{0in}\textbf{Proof.} }
\newcommand{\finishproof}{\hfill $\blacksquare$ \\}

\def\R{\mathbb{R}}

\newcommand{\so}{\mathfrak{so}}

\newcommand{\eqa}{\begin{eqnarray}}
\newcommand{\neqa}{\end{eqnarray}}
\newcommand{\be}{\begin{equation}}
\newcommand{\ee}{\end{equation}}

\newcommand{\scrM}{\mathcal{M}}

\newcommand{\scrD}{\mathcal{D}}

\newcommand{\scrZ}{\mathcal{Z}}
\newcommand{\scrL}{\mathcal{L}}
\newcommand{\scrK}{\mathcal{K}}
\newcommand{\scrV}{\mathcal{V}}
\newcommand{\scrO}{\mathcal{O}}
\newcommand{\scrH}{\mathcal{H}}
\newcommand{\scr}[1]{\mathcal{#1}}

\newcommand{\dual}{\,\,{}^\star\!}

\newcommand{\dif}{\mathrm{d}}

\newcommand{\double}[2]{\hspace{0.1em} #1 \hspace{#2} #1 \hspace{0.1em}}
\newcommand{\ident}{\double{1}{-0.35em}}

\newcommand{\pb}[1]{\underleftarrow{#1}}

\newcommand{\diag}{\mathrm{diag}}
\newcommand{\hateq}{\widehat{=}} 

\newcommand{\dummy}{\rule{0mm}{0mm}}

\newcommand{\topdec}[3]{\overset{\scriptscriptstyle #2}{\displaystyle #1}\rule{0pt}{#3}}
\newcommand{\gdec}[1]{\topdec{#1}{(\gamma)}{8pt}}

\newcommand{\toconsider}[1]{}

\newcommand{\sderiv}{\scrD}  

\begin{document}

\title{Purely geometric path integral for spin foams}
%
%

%
%
\author{
Atousa Chaharsough Shirazi\thanks{achahars@fau.edu} \; and \,
Jonathan Engle\thanks{jonathan.engle@fau.edu},\\
\\
{\it Department of Physics, Florida Atlantic University} \\
{\it 777 Glades Road, Boca Raton, FL 33431, USA }
}

\maketitle

%
%

\begin{abstract}

Spin-foams are a proposal for defining the dynamics of loop quantum gravity via path integral.  In order for a path integral to be at least formally equivalent to the corresponding canonical quantization, at each point in the space of histories it is important that the integrand have not only the correct phase -- a topic of recent focus in spin-foams -- but also the correct modulus, usually referred to as the measure factor.  The correct measure factor descends from the Liouville measure on the reduced phase space, and its calculation is a task of canonical analysis.

The covariant formulation of gravity from which spin-foams are derived is the Plebanski-Holst formulation, in which the basic variables are a Lorentz connection and a Lorentz-algebra valued two-form, called the Plebanski two-form. However, in the final spin-foam sum, one sums over only spins and intertwiners, which label eigenstates of the Plebanski two-form alone.  The spin-foam sum is therefore a discretized version of a Plebanski-Holst path integral in which only the Plebanski two-form appears, and in which the connection degrees of freedom have been integrated out. We call this a purely geometric Plebanski-Holst path integral.

In prior work in which one of the authors was involved, the measure factor for the Plebanski-Holst path integral with both connection and two-form variables was calculated.  Before one discretizes this measure and incorporates it into a spin-foam sum, however, one must integrate out the connection in order to obtain the purely geometric version of the path integral.  To calculate this purely geometric path integral is the principal task of the present paper, and it is done in two independent ways. Gauge-fixing and the background independence of the resulting path integral are discussed in the appendices.

\end{abstract}


\section{Introduction}

In the path integral approach to constructing a quantum theory, the integrand of the path integral
has two important parts: a phase part given by the exponential of $i$ times the classical action,
and a measure factor. The form of the phase part in terms of the classical action
ensures that solutions to the classical equations of motion dominate the path integral
in the classical limit so that one recovers classical physics in the appropriate regime.
The measure factor, however, arises from careful canonical analysis, and is important
for the path integral to be equivalent to the corresponding canonical quantum theory.
In most theories, this means that it is important, in particular, in order for the path integral theory to have such
elementary properties as yielding a unitary S-matrix that preserves probabilities.
The importance of having the correct measure factor is thus quite high.
%
%

Spin-foams are a path integral approach to quantum gravity in which one does not
sum over classical gravitational histories, but rather \textit{quantum} histories arising
from canonical quantization.  Specifically, in spin-foams, one sums over histories
of canonical states of \textit{loop quantum gravity}.  These histories possess a natural
4-dimensional space-time covariant interpretation, whence each can be thought of as
a \textit{quantum space-time}.
This approach allows one to retain the understanding gained from loop quantum
gravity, such as the discreteness of area and volume spectra, while at the same time formulating
the dynamics in a way that makes space-time symmetries more manifest \cite{rr1996}.

The starting point for the derivation of spin-foams is the \textit{Plebanski-Holst} formulation
of gravity \cite{epr2007, epr2007a, elpr2007, eht2009, engle2011}, in which the basic variables are a connection
$\omega$ and what is called the Plebanski two-form, $\Sigma$.
%
%
However, in the final spin-foam sum, the connection
$\omega$ is usually not present, and one sums over only spins and intertwiners, which
determine certain eigenstates of $\Sigma$ alone.  Because of this, the continuum
path integral most directly related to the spin-foam sum is a Plebanski-Holst path integral
in which only $\Sigma$ appears, and in which the connection has been integrated out.
We call such a path integral \textit{purely geometric} because $\Sigma$ directly determines the geometry
of space-time.

Because of the quantum mechanical nature of the histories used in spin-foams,
ensuring that the summand has the required phase part and measure factor is not completely trivial.
Only within the last couple years has the correct phase part been achieved \cite{engle2011a, engle2012}.
Regarding the measure factor, a first step has been carried out in the work \cite{eht2009}, where
the correct measure factor is calculated for the Plebanski-Holst path integral
with \textit{both} $\omega$ and $\Sigma$ present.
However, until now, the measure factor for the path integral with $\Sigma$ \textit{alone},
necessary for spin-foams, had not yet been calculated.  To carry out this calculation is the main purpose
of the present paper. In order to be sure about all numerical factors, we do this in two
different ways: (1.) by starting from the path integral in \cite{eht2009} and then integrating out
the connection degrees of freedom, and (2.) by starting from the ADM path integral and introducing
the necessary variables from there.  We find that these two ways of calculating the measure
factor exactly match, as must be the case, as the canonical measure factor ultimately descends from the
Liouville measure on the reduced phase space, which is independent of the formulation of gravity used \cite{eht2009}.

The path integral derived in this paper is ready to be discretized and translated into a spin-foam model, a task which will be carried out in forthcoming work.  Furthermore, when this is accomplished, we would like to emphasize that, because both primary and
secondary simplicity constraints are already incorporated in the continuum path integral \cite{eht2009, hs1994},
they will be automatically incorporated in the resulting spin-foam model as well.

It should also be kept in mind that the \textit{raison d'\^etre} of the canonical path integral is to ensure formal equivalence with canonical quantization, and it may be that one can use such equivalence as a more direct criterion for obtaining the correct `measure factor' in a spin-foam model.
The work \cite{brr2010} has begun to explore use of such alternative argumentation.
%
%

We begin the paper by reviewing some background on the path integral and how it is used.
After that, we derive the purely geometric Plebanski-Holst path integral, first starting from \cite{eht2009}, and then
starting from the ADM path integral.  
Gauge-fixing and the background independence of the derived path integral are then discussed in appendices.

\section{Background}
\label{bkgrsect}

\noindent\textit{Path integral generalities}

For an unconstrained field theory with canonically conjugate variables $\varphi, \pi$, the phase space path integral takes the
form\cite{weinberg1995}
\begin{equation}
\label{unconstrained}
\scrZ(\scrO) = \int \scrD \varphi \scrD \pi \scrO(\varphi,\pi) \exp i S[\varphi, \pi] .
\end{equation}
where the integration is over histories of pairs $(\varphi, \pi)$,
$S[\varphi,\pi]$ is the `phase space action'  $ \int \dif^4 x(\dot{\varphi} \pi - \scrH)$ with $\scrH$ the Hamiltonian density, and $\scrD \varphi \scrD \pi$ is a formal Cartesian product of Lesbesgue measures at each point in space-time
--- or, equivalently, a Cartesian product of \textit{Liouville measures} on phase space at each moment of time.
For a system with second class constraints, if $\bar\varphi,\bar\pi$ are taken to be canonically conjugate coordinates on the
\textit{constrained} phase space, then (\ref{unconstrained}) still applies.  However, if one uses
coordinates on the \textit{unconstrained} phase space, one
obtains\cite{ht1994}
\begin{equation}
\label{constr}
\scrZ(\scrO) = \int \scrD\varphi \scrD \pi \delta^n(C_i) |\det \{C_i, C_j\}|^{\frac{1}{2}}
\scrO \exp i S[\varphi, \pi]
\end{equation}
where the combined factor $\delta^n(C_i)|\{C_i, C_j\}|^{\frac{1}{2}}$
is independent of the way the constraint surface is represented by constraint functions $C_i$.
For a system with first class constraints $C_i$ in which the action takes the form
$S[\varphi,\pi,\lambda] = S_{o}[\varphi,\pi] + \int \lambda^i C_i$, with $\lambda^i$ Lagrange
multipliers,  one can write the path integral as \cite{fp1967}
\begin{equation}
\label{nogf}
\scrZ(\scrO) = \int \scrD \varphi \scrD \pi \scrD \lambda
\scrO \exp i S[\varphi,\pi,\lambda].
\end{equation}
Alternatively, one may introduce gauge fixing functions $\xi_i$ so that ${\xi}_i$ and $C_i$ together form a second class set of constraints, in which case (\ref{constr}) again applies.  If we assume that,
e.g., the $\xi_i$ are pure momentum or pure configuration, so that they
all Poisson commute, the path integral then takes the form \cite{faddeev1969}
\begin{equation}
\label{gf}
\scrZ(\scrO)
= \int \scrD \varphi \scrD \pi \scrD \lambda \delta^n(\xi)\, |\!\det \{C_i, \xi_j\}|\,
\scrO\exp i S[\varphi,\pi,\lambda] .
\end{equation}
In fact, under very general assumptions (which, however, have yet to be fully proven in the case of gravity)
(\ref{nogf}) and (\ref{gf}) are equal upto an infinite constant equal to the gauge
volume (see \cite{fp1967,han2009a, weinberg1995a}, and appendix \ref{equiv_app}), 
so that they determine the same physics in the manner to be reviewed below. 
In the rest of this paper for simplicity we will use expression (\ref{nogf}). However,
all derivations in this paper can just as well be done starting from expression (\ref{gf}) --
one need only reinsert the omitted factor
$\delta^n(\xi) |\det \{C_i, \xi_j\}|$.

The path integral was originally discovered as a way to write transition amplitudes between states in quantum mechanics.
Let $\{O_i\}$ denote a set of phase space functions whose quantum analogues
$\{\widehat{O}_i\}$ form a complete commuting set.
Let $\Psi_{\{(O_i, \nu_i)\}}$ denote the simultaneous eigenstate of the operators $\{\widehat{O}_i\}$
with eigenvalues $\{\nu_i\}$.
Then the path integral determines the transition amplitude between two such states via
\begin{eqnarray}
\left\langle \Psi_{\{(O_i', \nu_i')\}}, U(T'-T) \Psi_{\{(O_i, \nu_i)\}} \right\rangle
&=& \scrZ\left(\prod_{i} \delta(O_i'((\varphi, \pi)(T'))- \nu_i') \delta(O_i((\varphi,\pi)(T)) - \nu_i)\right)
\nonumber \\
&=& \int_{\substack{\\ \\ O_i((\varphi,\pi)(T)) = \nu_i \\  O_i'((\varphi,\pi)(T')) = \nu_i'  }}
\hspace{-0.5cm} \scrD \varphi \scrD \pi \scrD \lambda \delta^n(\xi)\, |\!\det \{C_i, \xi_j\}| \,
\exp i S[\varphi,\pi,\lambda] \hspace{1cm} \dummy
\label{trans}
\end{eqnarray}
where $U(T'-T)$ is the time evolution operator from $t=T$ to $t=T'$.
%
%
For a theory with first class constraints of which the Hamiltonian is a linear combination
--- i.e., a time reparametrization invariant theory ---
there is no time evolution operator, and, instead of (\ref{trans}), the interpretation of the path integral involves
a `rigging map' or `group averaging map' $\eta$ \cite{almmt1995} from kinematical (unconstrained) states to
physical states (states annihilated by the constraints) \cite{baez1994, rr1996, rovelli2004}:
\begin{equation}
\left\langle \eta\!\left(\Psi_{\{(O_i', \nu_i')\}}\right),
\eta\!\left(\Psi_{\{(O_i, \nu_i)\}}\right) \right\rangle_{phys}
= \int_{\substack{\\ \\ O_i((\varphi,\pi)(T)) = \nu_i \\  O_i'((\varphi,\pi)(T')) = \nu_i'  }}
\hspace{-0.5cm} \scrD \varphi \scrD \pi \scrD \lambda \delta^n(\xi)\, |\!\det \{C_i, \xi_j\}| \,
\exp i S[\varphi,\pi,\lambda] .
\label{tri_trans}
\end{equation}
This is the case relevant for us.
Note the physical inner product, and therefore $\scrZ(\scrO)$, is only physically meaningful up to rescaling by a constant.
%
%
In the following we will use the notation $\widehat{=}$ to denote ``equality upto rescaling by a constant.''

By setting $\scrO = 1$, one obtains the so-called partition function. Often in the literature one says that
the partition function defines the quantum theory.  Of course, what is meant is that the \textit{mathematical
form} --- or more precisely, the integrand --- of the partition function defines the quantum theory.
%
%
%
One then uses the integrand to integrate against general functions $\scrO$, the results of which are used as above.
In this paper, we are explicit about this fact for clarity.

The transition amplitude (\ref{trans}) or (\ref{tri_trans}) defines the \textit{dynamics} of the theory.
This, together with a \textit{kinematical} quantum framework of states and relevant operators, is enough to allow
one to calculate any quantity of physical interest.  In the case of GR, canonical loop quantum gravity provides
the kinematics.  In this case, by choosing the observables $O_i$ appropriately,
the canonical states $\Psi_{\{(O_i, \nu_i)\}}$ can be made to be
spin-network states \cite{rs1994}, generalized spin-network states \cite{al2004}, or Livine-Speziale
coherent states \cite{ls2007}.  In all of these cases, the relevant observables $O_i$
are purely geometric, depending only on the pull-back of $\Sigma_{\mu\nu}^{IJ}$ to the spatial slice.
%
%

\section{Derivation of a purely geometric path integral from Plebanski-Holst}

We will consider both Euclidean and Lorentzian gravity simultaneously, defining $s := +1$ and $s:=-1$ 
respectively in these two cases.  In additition to space-time manifold indices, denoted here by lower case Greek letters,
the Plebanski and Plebanski-Holst formulations of gravity make use of mixed tensors with `internal' indices
$I,J,K,\dots \in \{0,1,2,3\}$ which are raised and lowered with a fixed `internal' metric 
$\eta_{IJ}:= \diag(s,1,1,1)$.
The basic variables are
an $\so(\eta)$-valued connection $\omega_{\mu}^{IJ}$ and an $\so(\eta)$-valued two-form $X^{IJ}_{\mu\nu}$,
the Plebanski two-form.
In the theory, $X_{\mu\nu}^{IJ}$ is constrained to satisfy the so-called \textit{simplicity constraint}.
\begin{equation}
\label{simpconst}
C_{\mu\nu\rho\sigma} := \epsilon_{IJKL} X^{IJ}_{\mu\nu} X^{KL}_{\rho\sigma}
- \frac{s}{4!} \epsilon_{\mu\nu\rho\sigma} \epsilon^{\alpha\beta\gamma\delta}
\epsilon_{IJKL} X^{IJ}_{\alpha\beta} X^{KL}_{\gamma\delta} \approx 0
\end{equation}
where $\epsilon_{IJKL}$ denotes the `internal' alternating tensor.
This constraint has 20 independent components per point and restricts $X_{\mu\nu}^{IJ}$ to belong to one of five sectors.
The first is the degenerate sector,
in which $\epsilon^{\mu\nu\rho\sigma}\epsilon_{IJKL} X^{IJ}_{\mu\nu} X^{KL}_{\rho\sigma} = 0$.  The other four sectors
each correspond to $X_{\mu\nu}^{IJ}$ taking one of the following four forms 
\begin{displaymath}
\label{plebsectors}
X^{IJ} = \left\{\begin{array}{lc}
\pm \tfrac{1}{\kappa} e^I \wedge e^J & (I\pm) \\
\pm \tfrac{1}{2\kappa} \epsilon^{IJ}{}_{KL} e^K \wedge e^L & (II\pm)
 \end{array}\right.
\end{displaymath}
for some non-degenerate tetrad $e^I_\mu$ (see \cite{bhnr2004, eht2009}), where we have chosen to include the factor $\kappa:=16 \pi G$ here
instead of explicitly in front of the action (given below).
%
%
Given an element $Y^{IJ} \in \so(\eta)$, it is useful to
define
$\gdec{Y}^{IJ} := (Y - \frac{1}{\gamma} \dual Y)^{IJ} $, where $\gamma \in \R^+$ is the Barbero-Immirzi parameter,
and $(\dual Y)^{IJ} := \frac{1}{2}\epsilon^{IJ}{}_{KL} Y^{KL}$.  
%
%
Thus, in particular, $\gdec{X}^{IJ} := (X - \frac{1}{\gamma} \dual X)^{IJ}$.  Using these variables and notation, we start from
the Plebanski-Holst path integral derived in \cite{eht2009}
\begin{equation}
\label{plebpath}
\scrZ (\scrO) = \int_{(II\pm)} \scrD \omega_\mu^{IJ} \scrD X^{IJ}_{\mu\nu} \delta(C)
\scrV^9 V_s \scrO \exp i \int \gdec{X}_{IJ} \wedge F^{IJ}
\end{equation}
where the action is an integral over the space-time manifold $\scrM$,
and the ``$(II\pm)$'' subscript indicates that we are restricting
the integration to the $(II+)$ and $(II-)$ sectors, in which
%
%
\begin{equation}
\label{IIplus}
X^{IJ} = \pm \frac{1}{2\kappa} \epsilon^{IJ}{}_{KL} e^K \wedge e^L .
\end{equation}
The $\mathcal{V}$
appearing in (\ref{plebpath})
is the space-time volume density and $V_s$ is the spatial volume density determined by $e_\mu^I$.
For brevity, from now on
we will omit the ``$(II\pm)$'' subscript on the path integral (\ref{plebpath}), leaving it understood.
In the end, to ensure that solutions to Einstein's equations dominate in the classical
limit of the spin foam sum, it is necessary to restrict to a particular combination of sectors $(II+)$ and $(II-)$
called the \textit{Einstein-Hilbert} sector \cite{engle2011, engle2011a, engle2012}. However, this is not relevant for
the work of the present paper.
%
%
%
%
Without the simplicity constraint,
the action appearing in (\ref{plebpath}), with $B = \gdec{X}$,
is the so-called \textit{BF action}.
Restriction to the $(II\pm)$ sectors reduces
this action to the \textit{Holst} action \cite{holst1995}
\begin{displaymath}
S_{BF} \equiv \frac{1}{\kappa}\int (X_{IJ} - \frac{1}{\gamma} \dual X_{IJ}) \wedge F^{IJ} = \pm \frac{1}{2\kappa}\int (e_I\wedge e_J - \frac{1}{2\gamma} \epsilon_{IJKL} e^K \wedge e^L) \wedge F^{IJ} \equiv \pm S_{Holst}.
\end{displaymath}
%
%
The equations of motion derived by varying this action are
\begin{displaymath}
\dif_\omega \gdec{X}^{IJ} = 0 \; \Leftrightarrow \; \dif_\omega X^{IJ} = 0 \; \Leftrightarrow \; \dif_\omega e^I = 0,\quad \text{and} \quad e^\mu_I e^\nu_J F^{IJ}_{\mu\nu} = 0,
\end{displaymath}
the first being the torsion-free condition on $\omega$, and the second being equivalent to
the Einstein equations.

It should be noted that the path integral (\ref{plebpath}) is valid
only when $\scrO$ is purely geometric (i.e., a function of $X_{\mu\nu}^{IJ}$ only).
For, in the derivation in \cite{eht2009}, when the
Henneaux-Slavnov trick \cite{hs1994} is used, a change of variables
involving $\omega_\mu^{IJ}$ is performed.  If the presence of $\scrO$
is made explicit throughout the derivation, one sees that if $\scrO$ depends explicitly on
$\omega_\mu^{IJ}$, then the argument leading to (\ref{plebpath}) will change, as will the final
path integral expression\footnote{Specifically,
in the final path integral, if $\varphi_{HS}$ denotes
the Henneaux-Slavnov canonical transformation, then the $\omega$ argument of $\scrO$ will need to
be replaced by $\varphi_{HS} \cdot \omega$.}
However, the restriction of $\scrO$ to be purely geometric is not a real restriction, as,
on-shell, $\omega$ is uniquely determined by $X$. Furthermore, the principal application of (\ref{plebpath})
will be to computing the physical inner product between spin-network states, which are  eigenstates
of $X$ only (see equation (\ref{tri_trans})).

The goal of this section is to integrate out the connection in the expression (\ref{plebpath}). To make this
task easier, it is helpful to first show that the $\gamma$ dependent term
in the action can be dropped, which is done in the first of the following
subsections.  Using the fact that the integral over the connection is of Gaussian form,
in the second subsection we integrate out the connection, giving
the result in terms of the determinant of the appropriate matrix.  
The final
subsection computes the determinant of this matrix.

\subsection{Eliminating the $\gamma$-dependent term in the action}

Let
\begin{eqnarray*}
\scrL &:=& \frac{1}{3!} \epsilon^{\mu\nu\rho\sigma} X_{\mu\nu IJ} F^{IJ}_{\rho\sigma}\\
\gdec{\scrL} &:=& \frac{1}{3!} \epsilon^{\mu\nu\rho\sigma} \gdec{X}_{\mu\nu IJ} F^{IJ}_{\rho\sigma},
\end{eqnarray*}
the part of the Lagrangian in (\ref{plebpath}) without the $\gamma$ term, and
with the $\gamma$ term, respectively. For this subsection,
introduce a global time coordinate $t$, and let $\Sigma_{t'}$ denote the
hypersurface on which $t=t'$, providing us with a foliation of
$\scrM$ into hypersurfaces, and fix a vector field
$t^a$ such that $t^a \partial_a t = 1$.  We use lower case latin indices $a,b,c,...$ to denote indices associated with one or more of
the spatial manifolds $\Sigma_{t'}$ .
Let $\epsilon_{abc}$ and $\epsilon^{abc}$ denote the totally antisymmetric symbol
(i.e., Levi-Civita symbol of density weight $-1$ and $1$, respectively) on each hypersurface.
Then define
\begin{eqnarray}
\nonumber
G_{IJ} &:=& \epsilon^{abc} (\dif_\omega X)_{abc IJ}\\
\label{gaussdef}
\gdec{G}_{IJ} &:=& \epsilon^{abc} (\dif_\omega \gdec{X})_{abc IJ}.
\end{eqnarray}
Note that, if we define $\pi^a_{IJ}:=\epsilon^{abc} X_{bc IJ}$,
$\gdec{\pi}^a_{IJ}:=\epsilon^{abc} \gdec{X}_{bc IJ}$ and let
$\sderiv_a$ denote the derivative operator induced on each $\Sigma_t$
by the pull-back of $\omega$, then one sees that
$G_{IJ} = \sderiv_a \pi^a_{IJ}$ and $\gdec{G}_{IJ} := \sderiv_a \gdec{\pi}^a_{IJ}$
are the usual Gauss constraint without and with $\gamma$ term.
%
%
Finally, let
\begin{eqnarray*}
\scrK &:=& \scrL + \omega_t^{IJ} G_{IJ} = \epsilon^{abc} X_{ab}^{IJ} \partial_t \omega_{c IJ}
+ \epsilon^{abc} X_{ta}^{IJ} F_{bc IJ}\\
\gdec{\scrK} &:=& \gdec{\scrL} + \omega_t^{IJ} \gdec{G}_{IJ} =
\epsilon^{abc} \gdec{X}_{ab}^{IJ} \partial_t \omega_{c IJ}
+ \epsilon^{abc} \gdec{X}_{ta}^{IJ} F_{bc IJ} .
\end{eqnarray*}
The important point about these latter expressions are that
they are independent of $\omega_t^{IJ}$.
\begin{lemma}
\label{eomlemm}
$\gdec{G}_{IJ} = 0$ is equivalent to
$G_{IJ} = 0$, which is equivalent to
the 4-dimensional equation of motion
$\dif_\omega X^{IJ} = 0$.
\end{lemma}
{\startproof

\textit{(1.) Equivalence of $\gdec{G}_{IJ}=0$ and $G_{IJ}=0$:}

This is immediate from the fact that $G_{IJ}$ and $\gdec{G}_{IJ}$
are related by the map $Y^{IJ} \mapsto \gdec{Y}^{IJ}$, which is non-degenerate.\\
%
%

\textit{(2.) Equivalence of $G_{IJ} = 0$ and $\dif_\omega X^{IJ} = 0$:}

That $\dif_\omega X^{IJ} = 0$ implies $G^{IJ} = \epsilon^{abc} (\dif_\omega X)^{IJ}_{abc} = 0$ is immediate.
To show that $G^{IJ} = 0$ implies $\dif_\omega X^{IJ} = 0$,
we first decompose the latter into its pull-back $\pb{\dif_\omega X^{IJ}} = 0$
to each hypersurface and the pull-back of $t^a (\dif_\omega X^{IJ})_{abc} = 0$ to each hypersurface.
That $G^{IJ} = \epsilon^{abc}(\dif_\omega X)^{IJ}_{abc} = 0$ implies $\pb{\dif_\omega X^{IJ}} = 0$
is again immediate. Furthermore
\begin{eqnarray*}
\epsilon^{abc} t^d (\dif_\omega X)^{IJ}_{dab}
&=& \frac{1}{3!} \epsilon^{abc} t^d \epsilon_{dab} G^{IJ} \\
&=& \frac{1}{3} t^c G^{IJ}
\end{eqnarray*}
so that $G^{IJ}=0$ implies all of the components
of the equation $\dif_\omega X^{IJ} = 0$, as desired.
\finishproof}

\begin{lemma}
\label{Kcor}
The restriction to sectors $(II\pm)$ of the simplicity constraints and $G_{IJ} = 0$ together imply
\begin{displaymath}
\gdec{\scrK} = \scrK
\end{displaymath}
\end{lemma}
{\startproof
Because $G_{IJ} = 0$, from Lemma \ref{eomlemm} we have $\gdec{G}_{IJ} = 0$ and
$\dif_\omega X^{IJ} = 0$.  In sectors $(II\pm)$ of the simplicity constraints,
$X^{IJ} = \text{(const.)} \epsilon^{IJ}{}_{KL} e^K \wedge e^L$, so that $\dif_\omega X^{IJ} = 0$
becomes $\dif_\omega e^I = 0$, whose covariant exterior derivative
yields the Bianchi identity $F^I{}_J \wedge e^J = 0$,
which implies $\gdec{\scrL} = \scrL$.  Because $G_{IJ} = \gdec{G}_{IJ} = 0$,
this implies $\gdec{\scrK} = \scrK$.
\finishproof}

\begin{lemma}
\label{Gdelta_lemm}
\begin{equation}
\label{Gdeltas}
\delta^6(\gdec{G}_{IJ}) = \left|1-\frac{1}{\gamma^2}\right|^{-3} \delta^6(G_{IJ}).
\end{equation}
\end{lemma}
{\startproof
\begin{displaymath}
\delta^6(\gdec{G}_{IJ}) = \left|\det \frac{\partial \gdec{G}_{KL}}{\partial G_{MN}}\right|^{-1}
\delta^6(G_{IJ}).
\end{displaymath}
Decomposing $\gdec{G}$ and $G$ into self dual and anti-self-dual parts,
one is able to calculate the determinant on the right hand side, resulting in (\ref{Gdeltas}).
\finishproof}

\begin{proposition}
The $1/\gamma$ term in the partition function (\ref{plebpath})
can be dropped, so that
\begin{equation}
\label{nogamma}
\scrZ(\scrO) \hateq \int \scrD \omega_\mu^{IJ} \scrD X^{IJ}_{\mu\nu} \delta(C)
\scrV^9 V_s \scrO \exp i \int X_{IJ} \wedge F^{IJ} .
\end{equation}
\end{proposition}
{\startproof
\begin{eqnarray*}
\scrZ(\scrO) &=& \int \scrD \omega_\mu^{IJ} \scrD X^{IJ}_{\mu\nu} \delta(C)
\scrV^9 V_s \scrO \exp i \int \left(X_{IJ} - \frac{1}{\gamma} \dual X_{IJ} \right)\wedge F^{IJ} \\
&=& \int \scrD \omega_\mu^{IJ} \scrD X^{IJ}_{\mu\nu} \delta(C)
\scrV^9 V_s \scrO \exp i \int d^4 x (\gdec{\scrK} - \omega_t^{IJ} \gdec{G}_{IJ})\\
&=& \int \scrD \omega_a^{IJ} \scrD X^{IJ}_{\mu\nu} \delta(C) \delta(\gdec{G}_{IJ})
\scrV^9 V_s \scrO \exp i \int d^4 x \gdec{\scrK}\\
&\hateq& \int \scrD \omega_a^{IJ} \scrD X^{IJ}_{\mu\nu} \delta(C)
\delta(G_{IJ}) \scrV^9 V_s \scrO \exp i \int d^4 x \gdec{\scrK}\\
&\hateq& \int \scrD \omega_a^{IJ} \scrD X^{IJ}_{\mu\nu} \delta(C)
\delta(G_{IJ}) \scrV^9 V_s \scrO \exp i \int d^4 x \scrK\\
&\hateq& \int \scrD \omega_\mu^{IJ} \scrD X^{IJ}_{\mu\nu} \delta(C)
\scrV^9 V_s \scrO \exp i \int d^4 x (\scrK - \omega_t^{IJ}G_{IJ})\\
&\hateq& \int \scrD \omega_\mu^{IJ} \scrD X^{IJ}_{\mu\nu} \delta(C)
\scrV^9 V_s \scrO \exp i \int X_{IJ} \wedge F^{IJ}
\end{eqnarray*}
Where, in step 4, lemma \ref{Gdelta_lemm} has been used, and in step 5, lemma \ref{Kcor}
has been used.
\finishproof}

\subsection{Integrating out the connection}

The next step is to evaluate the Gaussian integral:
\begin{displaymath}
\label{gaussint1}
I(X) := \int \scrD \omega_\mu^{IJ} \exp i \int X_{IJ} \wedge F^{IJ}
= \int \scrD \omega_\mu^{IJ} \exp i \int X_{IJ} \wedge (\dif \omega^{IJ} + \omega^I{}_K \wedge \omega^{KJ}).
\end{displaymath}
Because, in the path integral (\ref{plebpath}), $X$ is constrained to be of the form (\ref{IIplus}),
we assume $X$ to be of this form throughout the section.
We begin by noting that if we define
\newcommand{\mata}[4]{a\Big[\rule{0in}{0.8em}^{#1}_{#3}\Big]\!\Big[\rule{0in}{0.8em}^{#2}_{#4}\Big]}
\begin{displaymath}
\mata{\alpha}{\beta}{IJ}{KL} := -\frac{1}{2} \epsilon^{\alpha\beta\gamma\rho}
X_{\gamma\rho}{}^{[I}{}_K \eta^{J]}{}_L 
%
%
\end{displaymath}
and 
\begin{displaymath}
b_{IJ}^{\alpha}:= \frac{1}{3!} \epsilon^{\alpha\beta\gamma\rho},
(\dif X_{IJ})_{\beta\gamma\rho}.
\end{displaymath}
the action becomes
\begin{displaymath}
S_{BF}[X, \omega] = \int d^4 x 
\left( \mata{\alpha}{\beta}{IJ}{KL} \omega_\alpha^{IJ} \omega_\beta^{KL} + b^\alpha_{IJ} \omega_\alpha^{IJ} \right)
\end{displaymath}
so that
\begin{displaymath}
I(X)
= \int \scrD \omega \exp i  \int d^4 x 
\left( \mata{\alpha}{\beta}{IJ}{KL}\omega_\alpha^{IJ} \omega_\beta^{KL} + b^\alpha_{IJ} \omega_\alpha^{IJ} \right).
\end{displaymath}
This casts the integral in the explicit Gaussian form in equations (\ref{gaussact}-\ref{gausspath}) in appendix \ref{gauss_app}.  
Using the result (\ref{gaussresult}), one has that, modulo an overall $X$-independent constant,
\begin{equation}
\label{finalI}
I(X) \hateq (\det a)^{-\frac{1}{2}} \exp i S[X]
\end{equation}
where $S[X] := S_{BF}[X, \omega[X]]$ with 
$\omega[X]$ the unique connection determined by $X$ via the
equation of motion of BF theory found by varying $\omega$,
\begin{displaymath}
\dif_{\omega[X]} X^{IJ} = \dif X^{IJ} + 2 \omega[X]^{[I}{}_K \wedge X^{|K| J]} = 0.
\end{displaymath}
%
%
Because $X$ is of the form (\ref{IIplus}), $\omega[X]$ is furthermore
the unique Lorentz spin-connection determined by the tetrad $e$, i.e.,
such that
\begin{displaymath}
\dif_\omega e^I = 0,
\end{displaymath}
so that the action reduces to the Holst and Palatini actions
\begin{displaymath}
S[X] \equiv S_{BF}[X, \omega[X]] \approx \pm S_{Holst}[e, \omega[e]]
= \pm S_{Palatini}[e, \omega[e]]
\end{displaymath}
where $\approx$ denotes equality when $X$ is of the form (\ref{IIplus}), 
and the last equality holds because, when $\omega = \omega[X]$, the extra `topological' term
added to the Palatini action to give the Holst action vanishes due to the Bianchi identity \cite{holst1995}.
%
%

In order to use equation (\ref{finalI}) in the path integral, it remains to calculate the determinant of $a$.

\subsection{The determinant of $a$}

For this calculation, it will be convenient to let
\begin{equation}
\label{rhodef}
\rho^{\mu}_{IJ} := \mata{\mu}{\nu}{IJ}{KL} \omega_\nu^{KL}
= \frac{1}{2} {\epsilon}^{\mu\beta\gamma\rho} X_{\gamma\rho}{}^{[I}{}_{K}
\omega_\beta^{J]K} .
\end{equation}
Define the inverse volume 4-form $V^{\alpha\beta\gamma\delta}:= \epsilon^{IJKL} e^\alpha_I e^\beta_J e^\gamma_K
e^\delta_L = \mathcal{V}^{-1} \epsilon^{\alpha\beta\gamma\delta}$.
Then
\begin{eqnarray}
\nonumber
\kappa \rho^{\mu IJ}
&=& \pm \frac{1}{4} {\epsilon}^{\mu\beta\gamma\rho}
\epsilon_{PKMN}e^M_\gamma e^N_\rho \eta^{PI} \omega_\beta^{JK}
- (I \leftrightarrow J) \\
\nonumber
&=& \pm \frac{1}{4} \mathcal{V} V^{\mu\beta\gamma\rho}
\epsilon_{PKMN}e^M_\gamma e^N_\rho \eta^{PI} \omega_\beta^{JK}
- (I \leftrightarrow J) \\
\nonumber
&=& \pm \frac{1}{4} \mathcal{V} e^{[\mu}_P e^{\beta]}_K \eta^{PI} \omega_\beta^{JK}
- (I \leftrightarrow J) \\
\nonumber
&=& \pm \frac{1}{4} \mathcal{V}  e^{[\beta}_K e^{\mu]I} \omega_\beta^{JK}
- (I \leftrightarrow J) \\
\label{omegarho_rel}
&=& \pm \frac{1}{2} \mathcal{V}  e^{[\beta}_{K} e^{\mu][I} \omega_\beta^{J]K}
\end{eqnarray}

Next,
we decompose the source ($\omega_\mu^{IJ}$) and target ($\rho^{\mu IJ}$), of the
matrix of interest $a$, in a way that will aide in the calculation.
First, we ``internalize'' the space-time indices:
\begin{equation}
\label{internal}
W_K{}^{IJ}:= e_K^\mu \omega_\mu^{IJ} \qquad P_K{}^{IJ} := e_{\mu K} \rho^{\mu IJ}
\end{equation}
and define
\begin{displaymath}
\begin{array}{l@{\qquad}l}
W^i_K := W_K{}^{0i}
& \tilde{W}^i_K := \frac{1}{2} \epsilon^i{}_{jl} W_K{}^{jl} \\
P^i_K := P_K{}^{0i}
& \tilde{P}^i_K := \frac{1}{2} \epsilon^i{}_{jl} P_K{}^{jl}
\end{array}
\end{displaymath}
and then decompose $W^i_K$ further into
\begin{eqnarray*}
W^i_0, && \\
W^i &:=& \frac{1}{2}\epsilon^{ij}{}_{k} W^k_j ,\\
\hat{W}^i_j &:=& \text{traceless part of }\frac{1}{2}(W^i_j + W^j_i), \text{and}
\\
W &:=& W^i_i.
\end{eqnarray*}
Similarly decompose $\tilde{W}^i_K$ into
$\tilde{W}^i_0, \tilde{W}^i, \hat{\tilde{W}}^i_j, \tilde{W}$, $P^i_K$ into $P^i_0, P^i, \hat{P}^i_j, P$,
and $\tilde{P}^i_K$ into
$\tilde{P}^i_0, \tilde{P}^i, \hat{\tilde{P}}^i_j, \tilde{P}$.
In terms of this decomposition of $\omega_\mu^{IJ}$ and $\rho^{\mu IJ}$,
(\ref{omegarho_rel}) becomes
\begin{equation}
\label{PWmatrix}
\begin{array}{l@{\qquad\quad}l}
\tilde{P}^i_0 = \mp \tfrac{s}{2\kappa} \scrV W^i
& P^i_0 = \mp \tfrac{1}{2\kappa} \scrV \tilde{W}^i \\
\tilde{P}^i = \pm \tfrac{1}{4\kappa} \scrV (\tilde{W}^i - W^i_0)
& P^i = \pm \tfrac{1}{4\kappa} \scrV (W^i - \tilde{W}^i_0) \\
\tilde{P} = \pm \tfrac{1}{2\kappa} \scrV \tilde{W}
& P = \pm \tfrac{1}{2\kappa} \scrV W \\
\hat{\tilde{P}}^i_j = \mp \tfrac{1}{4\kappa} \scrV \hat{\tilde{W}}^i_j
& \hat{P}^i_j = \mp \tfrac{1}{4\kappa} \scrV \hat{W}^i_j
\end{array}
\end{equation}

We are ready to calculate the determinant of $a$.  From (\ref{rhodef}) and (\ref{internal}),
\begin{equation}
\label{Adecomp}
\left|\det a\right| = \left|\det \left( \frac{\partial \rho^{\mu IJ}}{\partial \omega_\nu^{KL}}\right)\right|
= \left|\det \left( \frac{\partial \rho^{\mu IJ}}{\partial P_M{}^{KL}}\right)
\det \left( \frac{\partial P_M{}^{IJ}}{\partial W_N{}^{KL}}\right)
\det \left( \frac{\partial W_M{}^{IJ}}{\partial \omega_\mu^{KL}}\right)\right|
\end{equation}
The 24 by 24 middle matrix $\frac{\partial P_M{}^{IJ}}{\partial W_N{}^{KL}}$, expressed
in terms of the decompositions $P_M^{IJ} = (P^i_0, P^i, \hat{P}^i_j, P,
\tilde{P}^i_0, \tilde{P}^i, \hat{\tilde{P}}^i_j, \tilde{P})$
and $W_M^{IJ} = (W^i_0, W^i, \hat{W}^i_j, W, \tilde{W}^i_0, \tilde{W}^i, \hat{\tilde{W}}^i_j, \tilde{W})$,
from (\ref{PWmatrix}), is seen to be
\begin{equation}
\label{PWmat_expl}
\frac{\partial P_M{}^{IJ}}{\partial W_N{}^{KL}}
= \frac{\scrV}{4\kappa}\left(\begin{array}{cccccccc}
0 & 0 & 0 & 0 & 0 & -2 \cdot \ident_{3\times 3} & 0 & 0 \\
0 & \ident_{3\times 3} & 0 & 0 & -\ident_{3\times 3} & 0 & 0 & 0 \\
0 & 0 & -\ident_{5\times 5} & 0 & 0 & 0 & 0 & 0 \\
0 & 0 & 0 & 2 & 0 & 0 & 0 & 0 \\
0 & -2s\ident_{3\times 3} & 0 & 0 & 0 & 0 & 0 & 0 \\
-\ident_{3\times 3} & 0 & 0 & 0 & 0 & \ident_{3\times 3} & 0 & 0 \\
0 & 0 & 0 & 0 & 0 & 0 & -\ident_{5 \times 5} & 0 \\
0 & 0 & 0 & 0 & 0 & 0 & 0 & 2
\end{array}\right)
\end{equation}
where, recall, $s=+1$ for Euclidean gravity and $s=-1$ for Lorentzian gravity. This yields
\begin{displaymath}
\det \left(\frac{\partial P_M{}^{IJ}}{\partial W_N{}^{KL}}\right)
= \text{(const.)} \scrV^{24} .
\end{displaymath}
%
%
where the constant is just $(4\kappa)^{-24}$ time the determinant of the numerical matrix in (\ref{PWmat_expl}), which is non-zero, as one can check.
From (\ref{internal}),
\begin{displaymath}
\frac{\partial \rho^{\mu IJ}}{\partial P_M{}^{KL}}
= e^{\mu M} \delta^{[I}_{K} \delta^{J]}_L
\qquad
\frac{\partial W_M{}^{IJ}}{\partial \omega_\mu^{KL}}
= e^\mu_M \delta^{[I}_{K} \delta^{J]}_L
\end{displaymath}
which are both block diagonal in $[IJ], [KL]$.  Thus
\begin{displaymath}
\det\left(\frac{\partial \rho^{\mu IJ}}{\partial P_M{}^{KL}}\right)
= (\det e^{\mu I})^6 = (\det e^\mu_I)^6 = (\det e_\mu^I)^{-6} = \scrV^{-6}
\end{displaymath}
and
\begin{displaymath}
\det\left(\frac{\partial W_M{}^{IJ}}{\partial \omega_\mu^{KL}}\right)
= (\det e^\mu_I)^6 = \scrV^{-6}
\end{displaymath}
Substitution into (\ref{Adecomp}) finally gives
\begin{displaymath}
\left|\det a\right| = \text{(const.)} \scrV^{12} .
\end{displaymath}

\subsection{Final continuum path integral}

Using this expression in (\ref{finalI}) and substituting that into (\ref{nogamma})
yields the final pure geometric continuum path integral
\begin{equation}
\label{finalcontin}
\scrZ(\scrO) \hateq \int \scrD X^{IJ}_{\mu\nu} \delta(C)
\scrV^3 V_s \scrO \exp i \int X_{IJ} \wedge F[\omega[X]]^{IJ} .
\end{equation}
We wish to note that this expression derives from the full canonical
path integral, with primary as well as secondary simplicity constraints
imposed, and with the full requisite determinant factors.  The only assumption
necessary in the derivation is that $\scrO$ does not explicitly depend on $\omega$.
In spin foam sums, each spin foam is labeled by quantum numbers which are
a discrete analogue of precisely the two-form $X$,
making the above expression ideally suited for use with spin-foams.

\toconsider{Additionally, one could say, but I decided not to:
Even though the secondary constraints no longer explicitly appear,
one can see their explicit presence through ... (and then you might
come up with some argument).  But if you use any such argument,
its probably better at the quantum level.}

\section{ADM path integral}
\label{adm_sect}

As a secondary derivation of (\ref{finalcontin}) we start from the ADM formalism.
Because one has no second class constraints in the ADM formalism,
one doesn't need to use the Henneaux-Slavnov trick \cite{hs1994}.
As a consequence, this derivation can also be thought of as a `check' on the use of the Henneaux-Slavnov trick
in this case.

\subsection{Integrating out the momentum}

The non-gauge-fixed canonical path integral in the ADM formalism in terms of the canonical variables $(h_{ab},\pi^{ab})$ is
\begin{equation}
\label{eq:adm} \scrZ_{ADM}(\scrO)
= \int \mathcal{D}h_{ab}\mathcal{D}\pi ^{ab}\mathcal{D}N\mathcal{D}N^{a} \scrO \exp {i\int d^4x(\pi^{ab}\dot{h}_{ab}-\mathcal{H}_{ADM}[h,\pi,N,\vec{N}])}
\end{equation}
where recall $a,b,c\dots$ denote spatial indices.
Here $N$ and $N^a$ are lapse and shift lagrange multipliers and $\mathcal{H}_{ADM}$ is the Hamiltonian density given by \cite{wald1984}
\begin{displaymath}
\label{eq:ham}
\mathcal{H}_{ADM}=h^{\frac{1}{2}} N (-^{(3)}\textrm{R}+h^{-1}\pi^{ab}\pi_{ab}-\frac{1}{2}h^{-1}\pi^2) +2\pi^{ab}D_aN_b,
\end{displaymath}
where $h:=\det h_{ab}$, $\pi:={\pi^a}_a$, and $^{(3)}\textrm{R}$ denotes the scalar curvature of 
$h_{ab}$.
The integral (\ref{eq:adm}) can be written
\begin{equation}
\label{admgauss}
\scrZ_{ADM}(\scrO)=\int \mathcal{D}h_{ab}\mathcal{D}\pi ^{ab}\mathcal{D}N\mathcal{D}N^{a} \scrO \exp
{i\int d^4 x (A_{ab,cd}\pi^{ab}\pi^{cd}+B_{ab}\pi^{ab}+C)}
\end{equation}
where $A$, $B$ and $C$ are
\begin{displaymath}
A_{ab,cd}:=-\frac{Nh^{-\frac{1}{2}}}{2}(h_{ac}h_{bd}+h_{ad}h_{bc}-h_{ab}h_{cd}),
\end{displaymath}
\begin{displaymath}
B_{ab}:=\dot{h}_{ab}-2D_{(a}N_{b)}
\end{displaymath}
and
\begin{displaymath}
C:=Nh^{\frac{1}{2}}{^{(3)}\textrm{R}}.
\end{displaymath}
We assume $\scrO$ does not depend explicitly on $\pi$, so that (\ref{admgauss})
is again of the form (\ref{gaussact}-\ref{gausspath}) in appendix \ref{gauss_app},
and the integration over $\pi^{ab}$ yields
\begin{equation}
\label{admgaussres}
\scrZ_{ADM}(\scrO)\widehat{=}\int \mathcal{D}h_{ab}\mathcal{D}N\mathcal{D}N^{a} 
\left|\det A\right|^{-1/2}
\scrO \exp
{i\int d^4 x (\pi^{ab}[h, N, \vec{N}] \dot{h}_{ab} - \mathcal{H}_{ADM}[h, \pi[h, N, \vec{N}]])}
\end{equation}
where $\pi^{ab}[h, N, \vec{N}]$ is the value of $\pi^{ab}$ determined by $h_{ab}$, $N$, and $N^a$ via the appropriate equation of motion of the Hamiltonian theory.
Substituting the explicit expression for $\pi^{ab}[h,N,\vec{N}]$ into (\ref{admgaussres})  yields
\begin{equation}
\label{eq:admexponential}
\scrZ_{ADM}(\scrO) \widehat{=}\int \mathcal{D}h_{ab}\mathcal{D}N\mathcal{D}N^a  |\det A|^{-\frac{1}{2}} 
\scrO\exp i\int d^4x\mathcal{L}_{ADM}.
\end{equation}
where
\begin{displaymath}
\mathcal{L}_{ADM}:= N \sqrt{h} (K_{ab} K^{ab} - K^2 + ^{(3)}\mathrm{R})
\end{displaymath}
is the usual ADM Lagrangian with 
\begin{displaymath}
K_{ab} := \frac{1}{2} \mathcal{L}_n h_{ab} = \frac{1}{2N}(\dot{h}_{ab} - D_a N_b - D_b N_a)
\end{displaymath}
the usual extrinsic curvature.

It remains to find the determinant of $A$.

\subsection{The determinant of $A$}

By the symmetry of $h_{ab}$, there exists an orthogonal matrix ${O_{a}}^{e}$ such that ${O_{a}}^{e}{O_{b}}^{f}h_{ef}=\lambda_a\delta_{ab}$ for some $\lambda_a$ , so that
\begin{displaymath}
A'_{ab,cd}:= {O_{a}}^{e}{O_{b}}^{f}{O_{c}}^{g}{O_{d}}^{h}{A}_{ef,gh}=\frac{N}{2\sqrt{h}}[\lambda_a\lambda_b\delta_{ac}\delta_{bd}+\lambda_a\lambda_b\delta_{ad}\delta_{bc}
-\lambda_a\lambda_c\delta_{ab}\delta_{cd}].
\end{displaymath}
Now because rows $(ab)=(12)$, $(13)$ and $(23)$ in $A'$ each have just one non zero element ($A'_{12,12}$, $A'_{13,13}$ and $A'_{23,23}$, respectively) the determinant of the above equation yields
\begin{displaymath}
 A'_{12,12}A'_{13,13}A'_{23,23}  \det Q = \det A
\end{displaymath}
where $Q_{ab}:=A'_{aa,bb}$. Using $A'_{12,12}=\frac{N}{2\sqrt{h}}\lambda_{1} \lambda_{2}$ , $A'_{13,13}=\frac{N}{2\sqrt{h}}\lambda_{1} \lambda_{3}$ and
$A'_{23,23}=\frac{N}{2\sqrt{h}}\lambda_{2} \lambda_{3}$ gives
\begin{displaymath}
\det A=\left(\frac{N}{2\sqrt{h}}\right)^3(\lambda_1\lambda_2\lambda_3)^2 \det Q.
\end{displaymath}
Furthermore,
\begin{eqnarray*}
\det Q&=& \det \left( \begin{array}{ccc}
{A}_{11,11}&{A}_{11,22}&{A}_{11,33}\\
{A}_{22,11}&{A}_{22,22}&{A}_{22,33}\\
{A}_{33,11}&{A}_{33,22}&{A}_{33,33}\end{array}\right)
= \left(\frac{N}{2\sqrt{h}}\right)^3 \det \left( \begin{array}{ccc}
\lambda_{1}^{2} &-\lambda_1\lambda_2&\lambda_1\lambda_3\\
-\lambda_1\lambda_2&\lambda_{2}^{2}&-\lambda_2\lambda_3\\
-\lambda_1\lambda_3&-\lambda_2\lambda_3&\lambda_{3}^{2}\end{array}\right)\\
&=&- 4\left(\frac{N}{2\sqrt{h}}\right)^3(\lambda_1\lambda_2\lambda_3)^2
%
%
\end{eqnarray*}
So that,
\begin{displaymath} \det A=\left(\frac{N}{2\sqrt{h}}\right)^64(\lambda_1\lambda_2\lambda_3)^4=\frac{N^6}{2^4(\sqrt{h})^6}h^4=\frac{N^6}{2^4}h
\end{displaymath}
where $\lambda_1\lambda_2\lambda_3=\det h_{ab}$ has been used. Substituting this into
 (\ref{eq:admexponential})
gives
\begin{equation}
\label{eq:admfinal}
\scrZ_{ADM}(\scrO) \hateq
\int  \mathcal{D}h_{ab}\mathcal{D}N\mathcal{D}N^a N^{-3}h^{-\frac{1}{2}} \scrO \exp i\int d^4x\mathcal{L}_{ADM}.
\end{equation}

\subsection{Space-time covariant variables}

It remains to change from the variables $(h_{ab},N,N^{a})$ to the space-time covariant variable $g_{\mu\nu}$, we have
\begin{eqnarray*}
\mathcal{D}g_{\mu\nu}=\mathcal{D}g_{ab}\mathcal{D}g_{00}\mathcal{D}g_{0a}=\left|\det J\right| \mathcal{D}h_{ab}\mathcal{D}N\mathcal{D}N^a
\end{eqnarray*}
where
\begin{displaymath}
J:=\frac{\partial(g_{ab},g_{00},g_{0a})}{\partial(h_{cd},N,N^c)}.
\end{displaymath}
Using the relations
\begin{displaymath}
g_{ab}=h_{ab}  \qquad g_{00}=-N^2+h_{ab}N^aN^b \qquad g_{0a}=h_{ab}N^b
\end{displaymath}
%
%
it is easy to check that
\begin{displaymath}
\det J= -2hN
\end{displaymath}
so that
\begin{displaymath}
\mathcal{D}g_{\mu\nu}=2hN\mathcal{D}h_{ab}\mathcal{D}N\mathcal{D}N^a.
\end{displaymath}
Equation (\ref{eq:admfinal}) thus yields
\begin{equation}
\label{eq:measuremetric}\scrZ_{ADM}(\scrO)\hateq\int \mathcal{D}g_{\mu\nu} N^{-4}h^{-\frac{3}{2}}
\scrO \exp i S_{ADM}.
\end{equation}

\subsection{Comparison with the result of the last section}

Our goal in this section is to compare the ADM measure in (\ref{eq:measuremetric}) to the final continuum path integral (\ref{finalcontin}) derived in section 3
\begin{displaymath}
\label{finalcontinpartition}
\scrZ(\scrO) \hateq \int \scrD X^{IJ}_{\mu\nu} \delta(C)
\scrV^3 V_s \scrO \exp i \int X_{IJ} \wedge F[\omega[X]]^{IJ} .
\end{displaymath}
Since these two expressions have different variables, again a change of variables is necessary.
As discussed in \cite{eht2009}, one can perform a change of variables $X_{\mu\nu}^{IJ} \rightarrow ( e_{\mu}^I, C)$,
\footnote{The indices for $C$ are suppresed here for simplicity. In this paper $C$ is defined as (\ref{simpconst}) whereas in \cite{eht2009}
$C$ is defined in terms of three parts $(C^{ab},\tilde{C}_{\beta\beta},\tilde{C}_{\beta\pi})$. The formal Lebesgue measure and Dirac delta functions of these two formulations of the simplicity constraint are related by an irrelevant
constant Jacobian.}
%
%
from $X_{\mu\nu}^{IJ}$ to the tetrad $e^{I}_{\mu}$ together with the simplicity constraints.
The resulting change in measure is given by \cite{eht2009}

\begin{displaymath}
\mathcal{D}X_{\mu\nu}^{IJ}\hateq\mathcal{V}^{-6}\mathcal{D}e_{\mu}^I\mathcal{D}C
\end{displaymath}
so that
\begin{displaymath}
\scrZ(\scrO) \hateq \int \mathcal{D}e_{\mu}^I\mathcal{D}C\delta(C)\mathcal{V}^{-3}V_s \scrO
\exp i\int X_{IJ}\wedge F[\omega[X]]^{IJ} .
\end{displaymath}
Next, we change from tetrad variables to metric variables. This can be done by fixing an arbitrary reference tetrad
$\mathring{e}{}^{I}_{\mu}[g]$ for each metric $g_{\mu\nu}$, and defining a local gauge transformation ${\Lambda^{I}}_{J}$ via
\[e^{I}_{\mu}={\Lambda^{I}}_{J}\mathring{e}{}^{I}_{\mu}[g].\]
At each space-time point, ${\Lambda^{I}}_{J}$ is an element of either the 4-dimensional Euclidean group or the Lorentz group, depending on s. The change of variables $({e}_{\mu}^{I}) \rightarrow (g_{\mu\nu},{\Lambda_{I}}^{J})$ yields the change of measure \cite{acz2003}

\begin{displaymath}
\mathcal{D}g_{\mu\nu}\mathcal{D}{\Lambda_{I}}^J\hateq\sqrt{g} \mathcal{D}e_{\mu}^I
\end{displaymath}
 where $\mathcal{D}{\Lambda_{I}}^J$ is the measure defined in \cite{acz2003}. Because the integral
$\int \mathcal{D}\Lambda^I{}_J$ is independent of the metric, it can be dropped from the path integral, leaving
\begin{displaymath}
\scrZ(\scrO) \hateq \int\mathcal \mathcal{D}g_{\mu\nu}\mathcal{D}C\delta(C) {V}^{-4}V_s \scrO
\exp i\int X_{IJ}\wedge F[\omega[X]]^{IJ}.
\end{displaymath}
Using the relations ${V}_s=h^{\frac{1}{2}}$,  ${\mathcal{V}}=g^{\frac{1}{2}}=NV_s$
and the fact that $\int \mathcal{D}C\delta(C)=1$ gives
\begin{displaymath}
\scrZ(\scrO)\hateq \int \mathcal{D}g_{\mu\nu}   N^{-4} h^{-\frac{3}{2}} \scrO \exp i S_{ADM}
\end{displaymath}
which is the same as (\ref{eq:measuremetric}).

\section{Conclusions}

Spin-foams are a path integral approach to quantum gravity based on the Plebanski-Holst formulation
of general relativity.  The basic variables of the Plebanski-Holst formulation are a Lorentz connection and the Plebanski two-form, the pull-backs of which to any Cauchy surface are conjugate to each other.   The Plebanski two-form by itself completely
determines the space-time geometry.
In spin-foams, one sums over histories of spins and intertwiners
which label \textit{eigenstates of the Plebanski two-form}. Because of this, the spin-foam sum may be understood as
a discretization of a Plebanski-Holst path integral in which the connection
degrees of freedom have been integrated out --- that is, it is a discretization of what we have called a \textit{purely geometric}
Plebanski-Holst path integral.

In order to ensure that a path integral quantization be equivalent to canonical quantization,
it is important that the correct \textit{canonical path integral measure}
be used.  The path integral measure for Plebanski-Holst, with both connection and Plebanski two-form variables
present, was calculated in the earlier work \cite{eht2009}. In the present work, we have calculated the \textit{pure geometric}
form of this path integral, whose discretization will yield the necessary measure factor for spin-foams.
We have calculated the measure for this path integral in two independent ways (1.) by integrating out the connection
from the path integral derived in \cite{eht2009}, and (2.) by ensuring consistency with the canonical ADM path integral.
Both methods lead to the same final measure factor,
providing a check on the detailed powers of the space-time and spatial volume elements present.  The next step is to discretize
this measure on a spin-foam cell complex, expressing it directly in terms of spins and intertwiners.
This will involve non-trivial choices which will in part be fixed by considerations of gauge-invariance.  This will be discussed in
a later, complementary paper.

\section*{Acknowledgements}

The authors thank Thomas Thiemann and Christopher Beetle for discussions.  In particular,
JE is grateful to Thomas Thiemann for emphasizing the importance of including the correct
canonical measure factor in spin foams, and both authors thank Christopher Beetle for providing the
key ideas for the proof of lemma \ref{matrixlemm} in appendix \ref{equiv_app}.
This work was supported in part by the National Science Foundation through grant  PHY-1237510 and by
the National Aeronautics and Space Administration
through the University of Central Florida's NASA-Florida Space Grant Consortium.

\appendix

\section{Integration of a path integral with quadratic action}
\label{gauss_app}

Let $(\cdot,\cdot)$ be a symmetric, non-degenerate, but not necessarily positive definite, inner product on some real vector space
$V$.  Let $\hat{a}$ be an invertible operator on $V$ symmetric with respect to $(\cdot,\cdot)$, $b$ an element of $V$, and $c$ a real number.  Consider the action
\begin{equation}
\label{gaussact}
S[v]:= (v, \hat{a} v) + (v, b) + c 
\end{equation}
and the path integral 
\begin{equation}
\label{gausspath}
\int \mathcal{D} v e^{iS[v]}
\end{equation}
where $\mathcal{D} v$ is any translation-invariant measure on $V$ (unique up to rescaling).
If one fixes a basis on $V$ and uses components with respect to this basis as coordinates on $V$, 
$\mathcal{D}v$ will just be a real number times the Lebesgue measure in the chosen coordinates. From the complex analytic continuation of the usual formula for the Gaussian integral, one obtains
%
%
%
\begin{equation}
\label{gaussint}
\int \mathcal{D} v e^{iS[v]} = C
(\det \hat{a})^{-1/2} \exp \left( -\frac{i}{4}(b, \hat{a}^{-1} b) + c\right)
\end{equation}
%
%
where the constant $C$ depends exclusively on the relative scaling of the inner product $(\cdot,\cdot)$ and the measure $\mathcal{D} v$.
Variation of the action gives
\begin{displaymath}
\delta S|_{v=v_0} = (\delta v, \hat{a} v) + (v, \hat{a} \delta v) + (\delta v, b) 
= (\delta v, 2 \hat{a} v + b) .
\end{displaymath}
Setting this equal to zero for all $\delta v$ yields the following expression for the extremum $v_0$ of $S[v]$:
\begin{displaymath}
v_0 = - \frac{1}{2} \hat{a}^{-1} b.
\end{displaymath}
Substituting this into (\ref{gaussact}) directly gives $S[v_0] =  -\frac{i}{4}(b, \hat{a}^{-1} b) + c$, so that 
(\ref{gaussint}) can be written
\begin{equation}
\label{gaussresult}
\int \mathcal{D} v e^{iS[v]} = C (\det \hat{a})^{-1/2} e^{iS[v_0]}
\end{equation}
with $v_0$ the unique extremum of $S[v]$ and $C$ depending exclusively on the relative scaling of   $(\cdot,\cdot)$ and $\mathcal{D} v$.

\section{Background independence of the path integral}
\label{bi_app}

The path integral was originally discovered as a way to write transition
amplitudes in quantum mechanics.
%
%
In the case of a (time-)reparametrization invariant theory where
the Hamiltonian is a linear combination of the constraints (such as
GR), the path integral more precisely provides the projector onto physical states together with the
physical inner product \cite{rovelli2004, rr1996}.
As noted in section \ref{bkgrsect}, this physical inner product, together with the rest of the existing
canonical loop quantum gravity framework, is enough to calculate all physically
relevant quantities.
Let $\Sigma$ denote the spatial manifold of the canonical theory,
and let $\scrM$ denote a region of space-time with past
and future boundary $\Sigma_1, \Sigma_2$, each diffeomorphic
to $\Sigma$.  Select diffeomorphisms
$\iota_1 : \Sigma \rightarrow \Sigma_1 \subset \scrM$
and $\iota_2: \Sigma \rightarrow \Sigma_2 \subset \scrM$.
%
%
%
Select a set of observables $\{O_i\}$ which are functions of (the pull-back to $\Sigma$ of)
$X_{\mu\nu}^{IJ}$
only,
and whose quantum analogues $\{\widehat{O}_i\}$ form a complete communting set.
Let $\Psi_{\{(O_i, \nu_i)\}}$ denote the simultaneous eigenstate of $\{\widehat{O}_i\}$
satisfying $\widehat{O}_i \Psi_{\{(O_i, \nu_i)\}} =
\nu_i \Psi_{\{(O_i, \nu_i)\}}$.
As mentioned in section \ref{bkgrsect}, for different choices of $O_i$, these states could be spin networks \cite{rs1994}, generalized spin networks \cite{al2004},
or Livine-Speziale coherent states \cite{ls2007}.
From (\ref{tri_trans}), the path integral (\ref{finalcontin}) then determines the physical inner product
in LQG via
\begin{equation}
\label{physinner}
\langle \eta(\Psi_{\{(O_i', \nu_i')\}}), \eta(\Psi_{\{(O_i, \nu_i)\}}) \rangle_{phys}
\equiv \int_{\substack{O_i[\iota_1^* X] = \nu_i \\  O_i'[\iota_1^* X] = \nu_i'  }}
\mathcal{D}X_{\mu\nu}^{IJ} \mathcal{V}^{3} V_s \delta(C)e^{iS[X]} =: \scrZ[\Psi_{\{(O_i', \nu_i')\}}, \Psi_{\{(O_i, \nu_i)\}}]
\end{equation}
where $\eta$ is the `rigging map' or `group averaging map' from
kinematical (unconstrained) states to physical states
(states annihilated by the constraints)\footnote{In order
to construct such a map which also projects onto solutions of the Hamiltonian constraint,
one makes use of the Master constraint \cite{thiemann2003, dt2004}.
}, and $S[X]:= S_{BF}[X, \omega[X]]$ is the purely geometric action descending from BF theory.

The path integral measure in (\ref{physinner})
(as well as the path integral measures thus far derived in the literature for all other formulations of gravity \cite{bhnr2004, leutwyler1964, fv1973},
%
%
including those in equations  (\ref{plebpath}) and (\ref{eq:admfinal})) depends on (1.) a choice of foliation $\Xi$ of $\scrM$ (because of the
presence of 3-dimensional volume factors) and (2.)
a choice of coordinate system $\Phi$
%
%
compatible with $\Xi$ (because
the 4- and 3-volume factors are \textit{densities}).
That is, the measure depends on \textit{background} structures.
However, what matters physically is the physical inner product
in (\ref{physinner}), or more precisely, the physical inner product modulo
constant rescalings.
%
%
One can then ask: does the physical inner product (modulo rescalings)
determined by the above equation
retain this dependence on background?  In this section, we show that
the physical inner product is in fact background independent modulo rescalings, thus
respecting an important guiding principle from general relativity.

To begin the argument, we first note that any function can always be
made diffeomorphism covariant by making all background structure
an additional explicit argument.  Thus, if we express the transition amplitude
as a function of an initial state $\Psi_{\{(O_i, \nu_i)\}}$, a final state
$\Psi_{\{(O_i', \nu_i')\}}$, \textit{and} as a function of the
choice of foliation $\Xi$ and compatible coordinate system $\Phi$,
\begin{displaymath}
\scrZ[\Psi_{\{(O_i, \nu_i)\}}, \Psi_{\{(O_i', \nu_i')\}}, \Xi, \Phi],
\end{displaymath}
then it is by
construction diffeomorphism covariant. As $\scrZ$ is a scalar-valued
function, that means it is diffeomorphism \textit{invariant}:
\begin{equation}
\label{invariance}
\scrZ[\alpha \cdot \Psi_{\{(O_i, \nu_i)\}}, \alpha \cdot \Psi_{\{(O_i', \nu_i')\}},
\alpha \cdot \Xi, \alpha \cdot \Phi]
= \scrZ[\Psi_{\{(O_i, \nu_i)\}}, \Psi_{\{(O_i', \nu_i')\}}, \Xi, \Phi]
\end{equation}
for all diffeomorphisms $\alpha$ of $\scrM$.

Now, suppose $(\Xi, \Phi)$, $(\tilde{\Xi}, \tilde{\Phi})$ are two possible
choices of foliation and coordinate system.
%
%
Because the foliation arises from Feynman's procedure of skeletonization in time \cite{feynman1948}, the initial and
final slices $\Sigma_1$ and $\Sigma_2$ will always be leaves of $\Xi$ and $\tilde{\Xi}$.
%
%
Because of this, there always exists a 4-diffeomorphism $\alpha$
such that $\alpha \cdot \Xi = \tilde{\Xi}$, and such that $\alpha$ is the identity on
the initial and final hypersurfaces $\Sigma_1$ and $\Sigma_2$.
Because of the latter property,
$\alpha \cdot \Psi_{\{(O_i, \nu_i)\}} = \Psi_{\{(O_i, \nu_i)\}}$ and
$\alpha \cdot \Psi_{\{(O_i', \nu_i')\}} = \Psi_{\{(O_i', \nu_i')\}}$,
so that (\ref{invariance}) becomes
\begin{displaymath}
\scrZ[\Psi_{\{(O_i, \nu_i)\}}, \Psi_{\{(O_i', \nu_i')\}}, \tilde{\Xi}, \alpha \cdot \Phi]
= \scrZ[\Psi_{\{(O_i, \nu_i)\}}, \Psi_{\{(O_i', \nu_i')\}}, \Xi, \Phi].
\end{displaymath}
Next, note that, under a change of coordinate system,
the Lesbesgue measure, 3-volume and 4-volume densities, and Dirac delta function
in (\ref{physinner}) change only by Jacobian
factors which are constant on the space of histories.
Because of this, the left hand side of the above equation
is equal to $\scrZ[\Psi_{\{(O_i, \nu_i)\}}, \Psi_{\{(O_i', \nu_i')\}}, \tilde{\Xi}, \tilde{\Phi}]$ upto a constant,
so that
\begin{equation}
\label{Zindep}
\scrZ[\Psi_{\{(O_i, \nu_i)\}}, \Psi_{\{(O_i', \nu_i')\}}, \tilde{\Xi}, \tilde{\Phi}]
= \left(\text{const.} \right) \scrZ[\Psi_{\{(O_i, \nu_i)\}}, \Psi_{\{(O_i', \nu_i')\}}, \Xi, \Phi].
\end{equation}
where the constant is independent of $\Psi_{\{(O_i, \nu_i)\}}$ and $\Psi_{\{(O_i', \nu_i')\}}'$.
Let $\left[ \langle \eta(\cdot), \eta(\cdot)\rangle_{phys}\right]_{\Phi, \Xi}$
denote the physical inner product, as determined by the path integral using
$\Xi, \Phi$, modulo constant rescalings.  (\ref{Zindep}) then tells us that
\begin{displaymath}
\left[ \langle \eta(\cdot), \eta(\cdot)\rangle_{phys}\right]_{\tilde{\Phi}, \tilde{\Xi}}
=
\left[ \langle \eta(\cdot), \eta(\cdot)\rangle_{phys}\right]_{\Phi, \Xi} .
\end{displaymath}
Consequently
 $\left[ \langle \eta(\cdot), \eta(\cdot)\rangle_{phys}\right]$
 is independent of $\Xi$ and $\Phi$, and
 hence \textit{background independent}, as claimed.


\section{Equivalence of gauge-fixed and non-gauge-fixed path integrals}
\label{equiv_app}

In this appendix, we address the equivalence of the path integrals (\ref{nogf}) and (\ref{gf}).
The argument used here is based on the proof for the Yang-Mills case given by Faddeev and Popov \cite{fp1967}.
A version of this argument is also given in \cite{weinberg1995a} and \cite{han2009a}.
However, here we keep the argument more general and give more details.

\subsection{The argument}

Consider a system with first class constraints
$C_i$, generating a gauge group $\scr{G}$ on shell, and an action of the form
$S[\zeta, \lambda]= S_{o}[\zeta] + \int \lambda^i C_i$, where
$\zeta$ is short hand for a set of canonically conjugate variables $(\varphi,\pi)$, and
$\lambda$ are Lagrange multipliers.  We start with (\ref{nogf})
\begin{equation}
\label{nogf_app}
\scrZ(\scrO) = \int \scrD \zeta \scrD \lambda \scrO[\zeta]
\exp i S[\zeta ,\lambda]
= \int \scrD \zeta \delta(C(\zeta)) \scrO[\zeta]
\exp i S_o[\zeta],
\end{equation}
where $\scrD \zeta := \scrD \varphi \scrD \pi$, and where, for this appendix, we assume $\scrO$ is gauge invariant.
Faddeev and Popov in their original paper \cite{fp1967} almost start with this same path integral,
the only difference being that here we use phase space variables, which is the more fundamental
starting point from the canonical perspective \cite{faddeev1969, ht1994}.
%
%
%
The Faddeev-Popov strategy \cite{fp1967} is to factor out the divergences in the path integral
due to the integration over the gauge group.  We here adapt their argument to the general path integral
(\ref{nogf_app}) as follows.
First choose gauge-fixing functions $\xi_j = \xi_j(\zeta)$ which are regular, that is, have non-vanishing gradient,
at $\xi_j \equiv 0$. The $\xi_j$ then form a good set of coordinates
on each gauge-orbit in a neighborhood of $\xi_j \equiv 0$.  Furthermore, given any coordinates
$\alpha_i \mapsto g(\alpha) \in \mathcal{G}$ on the gauge group $\mathcal{G}$, and
any phase space point $\zeta = (\phi, \pi)$, one can define another set of coordinates on the gauge orbit containing $\zeta$ via $\alpha_i \mapsto g(\alpha) \cdot \zeta$.  One then has
\begin{displaymath}
\int \mathcal{D} \alpha \delta(\xi(g(\alpha)\cdot \zeta)) \left|\det \frac{\partial \xi^j(g(\alpha)\cdot \zeta)}{\partial \alpha^i}
\right|= \int \mathcal{D} \xi \delta(\xi) = 1
\end{displaymath}
which can be inserted into the path integral (\ref{nogf_app}) to obtain
\begin{equation}
\label{insertone}
\scrZ(\scrO) = \int \scrD \zeta \scrD \alpha \delta(C(\zeta))
\delta(\xi(g(\alpha)\cdot \zeta)) \left|\det \frac{\partial \xi_j(g(\alpha)\cdot \zeta)}{\partial \alpha^i}\right|\scrO[\zeta]
\exp i S_o[\zeta].
\end{equation}

We next perform the change of variables $\zeta \mapsto \zeta':= g(\alpha)\cdot \zeta$.
As $\zeta \mapsto g(\alpha)\cdot \zeta$ is a canonical transformation, and $d \zeta = d\varphi d\pi$ is the
Liouville measure, we have $d \zeta = d \zeta'$.
Furthermore, $S_o[\zeta] = S_o[\zeta']$ and $\mathcal{O}[\zeta] = \mathcal{O}[\zeta']$
as the action (and hence the action with constraints satisfied, $S_o$) and $\mathcal{O}[\zeta]$ are gauge-invariant.
%
%
It remains to consider the constraint factor $\delta(C(\zeta))$, and the determinant factor
$\left|\det \frac{\partial \xi^j(g(\alpha)\cdot \zeta)}{\partial \alpha^i}\right|$.
In order to facilitate calculation,  we now make a specific choice for $g(\alpha)$:
For each $\alpha^i$, we define $g(\alpha): \Gamma \rightarrow \Gamma$ to be the Hamiltonian flow generated
by $\alpha^i C_i$, evaluated at unit parameter time.
Equivalently, $g(\alpha)$ may be defined by the equationS
\begin{eqnarray}
\label{pbexp}
\{\alpha \cdot C, f\}|_{\tilde{\zeta}} &= &\left.\frac{\dif}{\dif t} f(g(-t\alpha)\cdot \tilde{\zeta})\right|_{t=0} ,\\
\nonumber
g(\vec{0})&=&\text{id}
\end{eqnarray}
%
%
If we let $X_{\alpha \cdot C}$ denote the derivative operator $X_{\alpha\cdot C} f := \{\alpha \cdot C, f\}$, then
the equations above implies the explicit expression
\begin{displaymath}
f(g(-\alpha) \cdot \tilde{\zeta}) = \sum_{n=0}^\infty \frac{1}{n!} \left( X_{\alpha\cdot C}\right)^n f |_{\tilde{\zeta}}
\equiv \left.\exp\left(X_{\alpha\cdot C}\right)f\right|_{\tilde{\zeta}}.
\end{displaymath}
The constraint factor $\delta(C(\zeta))$ can be rewritten $\delta(g(-\alpha)^*C(\zeta'))$,
where ${}^*$ denotes pull-back.
Because the flow generated by first class constraints is always tangent to the constraint surface,
$g(-\alpha)^* C_i$ will again be a linear combination of the constraints, so that
\begin{displaymath}
\label{transc}
g(-\alpha)^* C_i =: \mu(\alpha)_i{}^j C_j
\end{displaymath}
for some matrix $\mu(\alpha)_i{}^j$ of functions on phase space, whence
\begin{equation}
\label{deltatrans}
\delta(g(-\alpha)^* C(\zeta')) = |\det \mu(\alpha)|^{-1} \delta(C(\zeta')).
\end{equation}
Turning now to the last remaining factor, we have
\begin{eqnarray}
\nonumber
\frac{\partial \xi_j(g(\alpha)\cdot \zeta)}{\partial \alpha^i}
&=&  \left.\frac{d}{dt} \xi_j(g(\alpha + t \delta_i) \cdot \zeta) \right|_{t=0}
= \left.\frac{d}{dt} \xi_j(g(\alpha + t \delta_i) g(-\alpha) \cdot \zeta') \right|_{t=0} \\
\nonumber
&=& \int_0^1 ds \left.\frac{d}{dt} \xi_j(g(s\alpha)g(t \delta_i) g(-s\alpha) \cdot \zeta') \right|_{t=0}  \\
\nonumber
&=& \int_0^1 ds \left.\frac{d}{dt} (g(s\alpha)^*\xi_j)(g(t \delta_i) g(-s\alpha) \cdot \zeta') \right|_{t=0}
= \int_0^1 ds \{ C_i, g(s\alpha)^*\xi_j\}|_{g(-s\alpha) \cdot \zeta'} \\
\label{pdform}
&=&  \int_0^1 ds \{g(-s\alpha)^* C_i, \xi_j\}|_{\zeta'}
=  \left(\int_0^1 ds \mu(s\alpha)_i{}^k\right)\{C_k, \xi_j\}|_{\zeta'}
\end{eqnarray}
where $(\delta_i)^j:= \delta_i^j$.
Here the definition of the partial derivative has been used in the first line,
equation (\ref{nonabellemm}) from the next subsection has been used in the second line,
equation (\ref{pbexp}) has been used in the third line,
and the invariance of the Poisson bracket under the gauge transformation $g(s\alpha)$ has been used in the last line.
Define $M(\alpha)_i{}^j := \int_0^1 ds \mu(s\alpha)_i{}^j$.
Taking the determinant of (\ref{pdform}) then yields
\begin{displaymath}
\det \left(\frac{\partial \xi_j(g(\alpha)\cdot \zeta)}{\partial \alpha^i} \right)
= \det M(\alpha)|_{\zeta'} \; \det \{C_i, \xi_j\}|_{\zeta'}.
\end{displaymath}
Using this and equation (\ref{deltatrans}) in equation (\ref{insertone}) gives us
\begin{eqnarray*}
\scrZ(\scrO) &=& \int \scrD \zeta' \left(\int \scrD \alpha
\left|\frac{\det M(\alpha)}{\det \mu(\alpha)}\right|_{\zeta'} \right)
\delta(C(\zeta')) \delta(\xi(\zeta')) |\det \{C_i, \xi_j\}|_{\zeta'}
\scrO[\zeta'] \exp i S_o[\zeta'] \\
 &=& \int \scrD \zeta' \scrD \lambda \left(\int \scrD \alpha
\left|\frac{\det M(\alpha)}{\det \mu(\alpha)}\right|_{\zeta'} \right)
\delta(\xi(\zeta')) |\det \{C_i, \xi_j\}|_{\zeta'}
\scrO[\zeta'] \exp i S[\zeta', \lambda] .
\end{eqnarray*}
At this point, all dependence on $\alpha$ is restricted to the inner integral in parentheses, which can be thought of
as a ``gauge orbit volume,'' with
$\left|\frac{\det M(\alpha)}{\det \mu(\alpha)}\right|_{\zeta'}$
acting as a ``volume element''.
If we can show that this gauge orbit volume is independent of $\zeta'$, then we can
drop it as an overall constant, thereby proving the equivalence of the non-gauge-fixed (\ref{nogf_app}, \ref{nogf}) and
gauge-fixed (\ref{gf}) path integrals.

In the case of gravity, or any other theory with non-compact gauge orbits, this gauge orbit
volume is infinite, so it is not clear what it means to be constant on phase space.  In a moment, we will show that, in the case when
the algebra of constraints under consideration has structure \textit{constants},
then at least the gauge orbit volume \textit{element} is constant on phase space. Assuming that the ranges of the coordinates $\alpha^i$ on each gauge orbit are then also constant, this implies that the fully integrated gauge orbit volume itself is also constant, as required for equivalence.
%
%
%
%
%

The proof of the constancy of the gauge orbit volume element, $\left|\frac{\det M(\alpha)}{\det \mu(\alpha)}\right|$ , when there are structure constants,
is straight forward. We have
\begin{displaymath}
\mu(s\alpha)_i{}^j C_j := g(-s\alpha)^* C_i
=  \sum_{n=0}^{\infty} \frac{s^n}{n!} \{ \alpha \cdot C, \{ \alpha \cdot C, \dots \{ \alpha\cdot C, C_i\} \dots \}, \}
\end{displaymath}
where in each term there are $n$ nested Poisson brackets, with the $n=0$ term being just $C_i$.
If one has structure constants, each Poisson bracket introduces multiplication by a matrix which is
constant on phase space.  The product of these matrices, summed over $n$, is then equal to the matrix
$\mu(s\alpha)_i{}^j$ on the left hand side,
which is therefore also constant on phase space, leading to
$M(\alpha)_i{}^j$, and hence also $\left|\frac{\det M(\alpha)}{\det \mu(\alpha)}\right|$
being constant on phase space.

To handle the case of structure functions, which is in the case relevant for gravity,
one must look not only at the `gauge orbit volume element', but also at the fully integrated
`gauge orbit volume'.
As this volume is infinite, as already mentioned, it is not so clear what it means for it to be constant on the
phase space. A better understanding of how to handle this infinite volume through an appropriate
regulator, or experimentation with toy models with structure functions in which the gauge volume is finite,
could shed light on how to extend the last step of this proof of equivalence to systems of interest with structure functions.

\subsection{A result for non-Abelian gauge groups}

In this subsection we prove equation (\ref{nonabellemm}) below, which has been key in the above
subsection.  We begin by proving a general result for linear operators, which we then apply to the relevant case at hand.

\begin{lemma}
\label{matrixlemm}
For any two linear operators $A$ and $B$ on any vector space $V$,
\begin{displaymath}
\left.\frac{d}{dt} e^A e^{-tB-A}\right|_{t=0}
= \int_0^1 ds e^{sA} (-B) e^{-sA} .
\end{displaymath}
\end{lemma}
{\startproof
Let $I:= \left .\frac{d}{dt}[\exp A\exp (-tB-A)] \right |_{t=0}$. We then have
\begin{displaymath}
I=\int^{1}_{0}ds\left . \frac{\partial^2}{\partial{s}\partial{t}}[s\exp (sA) \exp (-tB-sA)] \right |_{t=0}
=\int^{1}_{0}ds\left . \frac{\partial^2}{\partial{s}\partial{t}}\left[s\exp (sA) \sum_{n=0}^{\infty}
\frac{(-tB-sA)^n}{n!}\right] \right |_{t=0} .
\end{displaymath}
Using the Leibniz rule to evaluate the derivative with respect to $t$, one obtains
\begin{eqnarray*}
I&=&\int^{1}_{0}ds \frac{\partial}{\partial{s}}\left[s\exp (sA) \sum_{n=0}^{\infty}\sum_{m=1}^{n}
\frac{1}{n!}(-sA)^{n-m}(-B)(-sA)^{m-1}\right]\\
&=&\int^{1}_{0}ds \frac{\partial}{\partial{s}}\left[\exp (sA)
\sum_{n=0}^{\infty}\sum_{m=1}^{n} \frac{(-s)^n}{n!}A^{n-m} B A^{m-1}\right]\\
&=&\int^{1}_{0}ds \left[\exp (sA)\sum_{m=1}^{\infty}\sum_{n=m}^{\infty}\left( \frac{(-s)^n}{n!}
A A^{n-m} B A^{m-1}+\frac{(-1)^n s^{n-1}}{(n-1)!}A^{n-m} B A^{m-1}\right)\right].
\end{eqnarray*}
Notice that we have switched the order of summation in the last line and used the fact that $A\exp(sA)=\exp(sA)A$.
Simplifying further,
\begin{displaymath}
I=\int^{1}_{0}ds \left[\exp (sA)\sum_{m=1}^{\infty}
\sum_{n=m}^{\infty}\left( \frac{(-s)^n}{n!} A^{n-m+1} B A^{m-1}
+\frac{-(-s)^{n-1}}{(n-1)!}A^{n-m} B A^{m-1}\right)\right].
\end{displaymath}
In this expression, the sum over $n$ telescopes, so that all terms in the sum cancel except the second
term in the $n=m$ case. One is thus left with
\begin{displaymath}
I=\int^{1}_{0}ds \left[\exp (sA)\sum_{m=1}^{\infty}\frac{-(-s)^{m-1}}{(m-1)!}B A^{m-1}\right]
=-\int^{1}_{0}\exp (sA) B \exp(-sA)ds.
\end{displaymath}
\finishproof}

\begin{proposition}
\begin{equation}
\label{nonabellemm}
\left.\frac{d}{dt} f(g(\alpha + t\beta) g(-\alpha) \zeta)\right|_{t=0}
= \int_0^1 ds \left.\frac{d}{dt}
f(g(s\alpha)g(t\beta)g(-s\alpha)\zeta)\right|_{t=0}
\end{equation}
\end{proposition}
{\startproof
\begin{eqnarray*}
\left.\frac{d}{dt} f(g(\alpha + t\beta) g(-\alpha) \zeta)\right|_{t=0}
&=& \left.\frac{d}{dt}\left[ \exp\left(X_{(-\alpha-t\beta)\cdot C}\right)f\right](g(-\alpha)\zeta)\right|_{t=0} \\
&=& \left.\frac{d}{dt}\left[ \exp\left(X_{\alpha\cdot C}\right)\exp\left(X_{(-\alpha-t\beta)\cdot C}\right)f\right]
(g(\zeta)\right|_{t=0}\\
&=& \left.\frac{d}{dt}\left[ \exp\left(X_{\alpha\cdot C}\right)\exp\left(-X_{\alpha\cdot C} -tX_{\beta\cdot C}\right)\right]_{t=0}
f\right|_\zeta \\
&=& \left.\int_0^1 ds
\left[ \exp\left(s X_{\alpha\cdot C}\right)(-X_{\beta\cdot C})\exp\left(-s X_{\alpha\cdot C}\right)\right]
f\right|_\zeta\\
&=& \left.\int_0^1 ds \left.\frac{d}{dt}
\left[ \exp\left(s X_{\alpha\cdot C}\right)\exp\left(-t X_{\beta\cdot C}\right)\exp\left(-s X_{\alpha\cdot C}\right)\right]\right|_{t=0}
f\right|_\zeta\\
&=& \int_0^1 ds \left.\frac{d}{dt}
\left[ \exp\left(X_{s\alpha\cdot C}\right)\exp\left(X_{-t\beta\cdot C}\right)\exp\left(X_{-s\alpha\cdot C}\right)f\right]
(\zeta)\right|_{t=0}\\
&=& \int_0^1 ds \left.\frac{d}{dt}
f(g(s\alpha)g(t\beta)g(-s\alpha)\zeta)\right|_{t=0}
\end{eqnarray*}
where, in the fourth line, lemma \ref{matrixlemm} was used in the case where $V$ is the space of functions on $\Gamma$,
$A = X_{\alpha \cdot C}$, and $B= X_{\beta \cdot C}$.
\finishproof}


\begin{thebibliography}{10}

\bibitem{rr1996}
M.~Reisenberger and C.~Rovelli, ````{S}um over surfaces'' form of loop quantum
  gravity,'' {\em Phys. Rev. D}, vol.~56, pp.~3490--3508, 1997.

\bibitem{epr2007}
J.~Engle, R.~Pereira, and C.~Rovelli, ``The loop-quantum-gravity
  vertex-amplitude,'' {\em Phys. Rev. Lett.}, vol.~99, p.~161301, 2007.

\bibitem{epr2007a}
J.~Engle, R.~Pereira, and C.~Rovelli, ``Flipped spinfoam vertex and loop
  gravity,'' {\em Nucl. Phys. B}, vol.~798, pp.~251--290, 2008.

\bibitem{elpr2007}
J.~Engle, E.~Livine, R.~Pereira, and C.~Rovelli, ``{L}{Q}{G} vertex with finite
  {I}mmirzi parameter,'' {\em Nucl. Phys. B}, vol.~799, pp.~136--149, 2008.

\bibitem{eht2009}
J.~Engle, M.~Han, and T.~Thiemann, ``Canonical path integral measures for
  {H}olst and {P}lebanski gravity. {I}. {R}educed phase space derivation,''
  {\em Class. Quant. Grav.}, vol.~27, p.~235024, 2009.

\bibitem{engle2011}
J.~Engle, ``{The Plebanski sectors of the EPRL vertex},'' {\em Class. Quant.
  Grav.}, vol.~28, p.~225003, 2011.
\newblock Corrigendum: {\it Class. Quant. Grav.} vol. 30, p. 049501, 2013.

\bibitem{engle2011a}
J.~Engle, ``A proposed proper {EPRL} vertex amplitude,'' {\em Phys. Rev. D},
  vol.~87, p.~084048, 2013.

\bibitem{engle2012}
J.~Engle, ``A spin-foam vertex amplitude with the correct semiclassical
  limit,'' {\em Phys. Lett. B}, vol.~724, pp.~333--337, 2013.

\bibitem{hs1994}
M.~Henneaux and S.~Slavnov, ``A note on the path integral for systems with
  primary and secondary second class constraints,'' {\em Phys. Lett. B},
  vol.~338, pp.~47--50, 1994.

\bibitem{brr2010}
E.~Bianchi, D.~Regoli, and C.~Rovelli, ``{Face amplitude of spinfoam quantum
  gravity},'' {\em Class.Quant.Grav.}, vol.~27, p.~185009, 2010.

\bibitem{weinberg1995}
S.~Weinberg, {\em The Quantum Theory of Fields. I. Foundations}.
\newblock Cambridge: Cambridge UP, 1995.

\bibitem{ht1994}
M.~Henneaux and C.~Teitelboim, {\em Quantization of Gauge Systems}.
\newblock Princeton: Princeton UP, 1994.

\bibitem{fp1967}
L.~Faddeev and V.~Popov, ``Feynman diagrams for the yang-mills field,'' {\em
  Phys. Lett. B}, vol.~25, pp.~29--30, 1967.

\bibitem{faddeev1969}
L.~Faddeev, ``{The Feynman integral for singular Lagrangians},'' {\em
  Theoretical and Mathematical Physics}, vol.~1, pp.~1--13, 1969.
\newblock Transl. of \textit{Teor. Mat. Fizika}, vol. 1, pp. 3-18, 1969.

\bibitem{han2009a}
M.~Han, ``{Canonical Path-Integral Measures for Holst and Plebanski Gravity.
  II. Gauge Invariance and Physical Inner Product},'' {\em Class.Quant.Grav.},
  vol.~27, p.~245015, 2010.

\bibitem{weinberg1995a}
S.~Weinberg, {\em The Quantum Theory of Fields. II. Modern Applications}.
\newblock Cambridge UP, 1995.

\bibitem{almmt1995}
A.~Ashtekar, J.~Lewandowski, D.~Marolf, J.~Mour\~ao, and T.~Thiemann,
  ``Quantization of diffeomorphism invariant theories of connections with local
  degrees of freedom,'' {\em J. Math. Phys.}, vol.~36, pp.~6456--6493, 1995.

\bibitem{baez1994}
J.~Baez, ``Strings, loops, knots and gauge fields,'' in {\em Knots and Quantum
  Gravity} (J.~Baez, ed.), Oxford: Oxford UP, 1994.

\bibitem{rovelli2004}
C.~Rovelli, {\em Quantum Gravity}.
\newblock Cambridge: Cambridge UP, 2004.

\bibitem{rs1994}
C.~Rovelli and L.~Smolin, ``Discreteness of area and volume in quantum
  gravity,'' {\em Nucl. Phys. B}, vol.~442, pp.~593--622, 1995.

\bibitem{al2004}
A.~Ashtekar and J.~Lewandowski, ``Background independent quantum gravity: A
  status report,'' {\em Class. Quant. Grav.}, vol.~21, p.~R53, 2004.

\bibitem{ls2007}
E.~Livine and S.~Speziale, ``A new spinfoam vertex for quantum gravity,'' {\em
  Phys. Rev. D}, vol.~76, p.~084028, 2007.

\bibitem{bhnr2004}
E.~Buffenoir, M.~Henneaux, K.~Noui, and P.~Roche, ``Hamiltonian analysis of
  {P}lebanski theory,'' {\em Class. Quant. Grav.}, vol.~21, pp.~5203--5220,
  2004.

\bibitem{holst1995}
S.~Holst, ``Barbero's {H}amiltonian derived from a generalized
  {H}ilbert-{P}alatini action,'' {\em Phys. Rev. D}, vol.~53, pp.~5966--5969,
  1996.

\bibitem{wald1984}
R.~M. Wald, {\em General Relativity}.
\newblock Chicago: Chicago University Press, 1984.

\bibitem{acz2003}
R.~Aros, M.~Contreras, and J.~Zanelli, ``{Path integral measure for first order
  and metric gravities},'' {\em Class.Quant.Grav.}, vol.~20, pp.~2937--2944,
  2003.

\bibitem{thiemann2003}
T.~Thiemann, ``The phoenix project: Master constraint programme for loop
  quantum gravity,'' {\em Class. Quant. Grav.}, vol.~23, pp.~2211--2248, 2006.

\bibitem{dt2004}
B.~Dittrich and T.~Thiemann, ``Testing the master constraint programme for loop
  quantum gravity {I}. {G}eneral framework,'' {\em Class. Quant. Grav.},
  vol.~23, pp.~1025--1066, 2006.

\bibitem{leutwyler1964}
H.~Leutwyler, ``{Gravitational field: Equivalence of Feynman quantization and
  canonical quantization},'' {\em Phys.Rev.}, vol.~134, pp.~B1155--B1182, 1964.

\bibitem{fv1973}
E.~Fradkin and G.~Vilkovisky, ``{S matrix for gravitational field. ii. local
  measure, general relations, elements of renormalization theory},'' {\em
  Phys.Rev.}, vol.~D8, pp.~4241--4285, 1973.

\bibitem{feynman1948}
R.~Feynmen, ``Space-time approach to non-relativistic quantum mechanics,'' {\em
  Rev. Mod. Phys.}, vol.~20, pp.~367--387, 1948.

\end{thebibliography}
%
%

\end{document}